\documentclass[12pt]{article}
\usepackage{amssymb}
\usepackage{latexsym}
\newcommand{\Zz}{{\mathbb Z}}

\newcommand{\Rr}{{\mathbb R}}

\def\be{\begin{equation}\label}
\def\ee{\end{equation}}
\def\bea{\begin{eqnarray}\samepage\label}
\def\eea{\end{eqnarray}}
\def\d{{\rm d}}
\def\a{{\bf a}}
\def\b{{\bf b}}
\def\k{{\bf k}}
\def\e{{\bf e}}
\def\kbar{{\overline{\k}}}
\def\kprimbar{{\overline{\k'}}}
\def\r{{\bf r}}
\def\p{{\bf p}}
\def\u{{\bf u}}
\def\v{{\bf v}}
\def\calp{{\mathcal P}}
\def\Pal{{\P^\alpha}}
\def\Pbet{{\P^\beta}}
\def\pal{\p^\alpha}
\def\pbet{\p^\beta}
\def\q{{\bf q}}
\def\n{{\bf n}}
\def\R{{\bf R}}
\def\K{{\bf K}}
\def\P{{\bf P}}
\def\Q{{\bf Q}}
\def\+{\!+\!}
\def\vol{|\Lambda|}
\def\calg{{\mathcal G}}
\begin{document}
\title{\large\bf Exact eigenstates for contact interactions
\footnote{Dedicated to P\'eter Sz\'epfalusy for his 70th and to Domokos Sz\'asz
for his 60th birthday.}
}
\author{Andr\'as S\"ut\H o\\
Research Institute for Solid State Physics and Optics\\
Hungarian Academy of Sciences\\
P. O. Box 49, H-1525 Budapest 114, Hungary\\
Email: suto@szfki.hu}
\date{}
\maketitle
\thispagestyle{empty}
\begin{abstract}
\noindent

We show that in $d\geq 2$ dimensions the $N$-particle kinetic energy operator with periodic boundary
conditions has symmetric eigenfunctions which
vanish at particle encounters, and give a full description of these functions.
In two and three dimensions they represent common eigenstates of bosonic Hamiltonians with any kind of
contact interactions, and
illustrate a partial `multi-dimensional Bethe Ansatz' or `quantum KAM theorem'.
The lattice analogs of these functions exist for $N\leq L^{[d/2]}$
where $L$ is the linear size of the box,
and are common eigenstates of Bose-Hubbard Hamiltonians
and spin-${1\over 2}$ XXZ Heisenberg models.
\vspace{2mm}
\end{abstract}
\newpage
\section{Introduction}
In this paper we construct eigenstates to Hamiltonians of the form
\bea{H}
&&H_N=-\sum_{i=1}^N\Delta_i+V(\r_1,\ldots,\r_N)-\mu N\nonumber\\
&&V(\r_1,\ldots,\r_N)=0 \mbox{ if all $\r_i$ are different,}
\eea
that is, Hamiltonians with real symmetric contact interactions.
The definition of such operators with the use of Dirac deltas is straightforward in one dimension and
leads to integrable systems \cite{LLY}. In higher dimensions the mere definition becomes mathematically
tricky. The two- and three-dimensional realization of (\ref{H}) needs renormalization to zero
of the interaction strength \cite{BF}, while in four dimensions and above
the family (\ref{H}) reduces to a unique element with $V\equiv 0$, i.e. the kinetic energy operator
(\cite {RS}, Theorem X.11).
One can gain an impression
about how this conclusion emerges by observing that the existence of contact interactions
is closely related to the existence of a Green's function for the Laplacian, i.e. an $L^2$-solution
$G_z(\cdot -\r')$ of the inhomogeneous equation
\be{Green}
(-\Delta_\r-z)G_z(\r-\r')=\delta(\r-\r')
\ee
for any complex (non-real) $z$. Any Hamiltonian with a contact interaction is a self-adjoint
extension of the N-particle Laplacian defined on $C^\infty$-functions whose support avoids the set
$\{\r_i=\r_j, i\neq j\}$. This symmetric operator
must have non-vanishing deficiency indices, i.e.
its adjoint must have eigenstates with complex eigenvalues. For two particles, these eigenstates are
the Green's functions for complex values of $z$.
If an $L^2$-solution exists for equation (\ref{Green}), it has a Fourier representation
\be{Fourier}
G_z(\r-\r')={1\over (2\pi)^d}\int{e^{i\p (\r-\r')}\over \p^2-z} \d\p\ .
\ee
In one dimension this integral is convergent for any $\r-\r'$ and yields a bounded continuous function.
In two and three dimensions $\int |\p^2-z|^{-2}\d\p$ is finite.
Therefore, by Plancherel's theorem, $(\p^2-z)^{-1}$
is the Fourier transform of an $L^2$-function which, however, diverges at $\r=\r'$ as
$\ln |\r-\r'|$ and $|\r-\r'|^{-1}$, respectively.
In four dimensions and above $\int |\p^2-z|^{-2}\d\p=\infty$
and, thus, (\ref{Green}) has no square-integrable solution.
Now by separating off the motion of the center of mass,
the two-body problem with a $\delta$-interaction can be reduced to a one-body problem in a
potential $\alpha\delta$. Multiplication of a suitable function $f$ with $\alpha\delta$ has to
be interpreted as $\alpha\langle\delta,f\rangle\delta$, where $\langle\delta,\cdot\rangle$ denotes
the action of the linear functional $\delta$. %and the prefactor of $\delta$ should be finite.
If we want to define $A=-\Delta +\alpha\langle\delta,\cdot\rangle\delta$ as a self-adjoint operator
on $L^2$($\Rr^d$), we have to restrict it to functions $\psi\in L^2$ such that $A\psi$ is also in
$L^2$ and $(\psi,A\psi)$ is real. So $A\psi$ cannot contain a $\delta$, and this is why $G_z$ has to be
included in $\psi$. Actually $z=i$ suffices, and a straightforward computation yields that a general
$\psi$ in the domain of $A$ is of the form
\be{domain}
\psi(\r)=\tilde{\psi}(\r)+a\tilde{\psi}(0)\,{\rm Re}\,G_i(\r)
\ee
with
\be{a}
a=-{\alpha\over 1+\alpha\langle\delta,{\rm Re}\,G_i\rangle}
\ee
and then
\be{Apsi}
A\psi=-\Delta\tilde\psi+a\tilde{\psi}(0)\,{\rm Re}\,iG_i\ .
\ee
Here $\tilde{\psi}$ is any vector from the domain of the kinetic energy $-\Delta$. The problem appears in the
denominator of $a$, with $\langle\delta,{\rm Re}\,G_i\rangle$.
This is well-defined in one dimension and yields Re$\,G_i(0)={1\over 2\sqrt{2}}$.
In dimensions two and three $\delta$ is not defined on Re$\,G_i$. One can proceed at least
in two different ways. One may define $\delta$ so that it does not feel the singularity of $G_i$ and set
$\langle\delta,{\rm Re}\,G_i\rangle=c$. Then
different constants yield  different operators, e.g $c=\infty$ gives $A=-\Delta$. This more
recent method was followed in \cite{AlK}. The original procedure of \cite{BF} replaced $\delta$ by a
sequence $\delta_n$ of regular approximants and $\alpha$ by $\alpha_n$. When $\delta_n\to\delta$,
$\langle\delta_n,{\rm Re}\,G_i\rangle\to\infty$. In order to obtain a nonvanishing limit for $a$, $\alpha_n$
then has to go to zero; otherwise we get back the kinetic energy operator.

The general, $N$-particle case is much more involved since it requires considering multiple
collisions and operators with infinite deficiency indices. The crucial question is to find the
physically meaningful (lower semibounded) self-adjoint extensions. For this purpose generalized
contact interactions have to be introduced \cite{KP}, \cite{AK}.
For an exhaustive mathematical treatment the reader can consult the monographs
\cite{AGHH} and \cite{AK} which also contain a detailed bibliography
of the abundant mathematical and physical literature. As an example from the latter, we mention the use of
contact interactions as `Fermi pseudopotentials' for an approximate treatment of the hard-core Bose gas
\cite{HYLLW}. They also appear in connection with the Laughlin wavefunction \cite{TK} and the trapped
boson gas \cite{BP}, \cite{WGS}, in which case the ground state of the projected
(to the lowest oscillator level) Hamiltonian for fixed nonvanishing angular momenta
was found. Projection to the lowest oscillator level has a regularizing effect on the interaction, which is
however not sufficient to settle all problems of definition.
To avoid the lower unboundedness of the
projected Hamiltonian and to make sure that the ground state energy does not {\it decrease} with an
increasing angular momentum,
the interaction strength still has to decrease as $1/N$ with the number of particles, see equations (4) and
(5) of Ref. \cite{BP}.
We note that apparently all existing
mathematical results concern systems of particles
in infinite space, and cannot but anticipate the extension of these results to confined systems.

We are interested in the construction of eigenstates to the Hamiltonians (\ref{H})
for the following reason.
The delta-gas in one dimension is an integrable system, and we would like to see
whether in two and three dimensions it preserves some trace of integrability.
A partial survival of integrability could be
viewed either as a remnant of the Bethe Ansatz when the dimensionality is
increased or as a germ of a
quantum-KAM theorem when, in the free Bose-gas, a contact interaction is 
switched on. The Bethe Ansatz
solution of the one-dimensional delta-gas assigns 
$N$ wave numbers to each eigenfunction. The two
and three dimensional analogy would be the existence of eigenstates which 
can be expressed in terms of finitely many wave vectors.
From the point of view of the KAM theory, surviving, maybe slightly distorted,
free-particle eigenstates would correspond to
classical KAM tori \cite{Kunz}. 
What we actually find are eigenstates of $N$ bosons in a box of
linear size $L$ which remain unperturbed by contact interactions, are
characterized by $2N$ wave vectors and
have an energy at least of order $N^3/L^2$.

A common feature of the operators (\ref{H}) is that, on smooth functions which
vanish at particle encounters, all act
as the pure kinetic energy operator. Therefore, if
this latter has eigenstates vanishing whenever $\r_i=\r_j$ for some $i\neq j$,
these will be common eigenstates of all members of the family (\ref{H}).
For two particles in two dimensions it is very easy to give examples.
Probably the simplest of them are
$$\sin{2n\pi\over L}(x_1-x_2)\,\sin{2m\pi\over L}(y_1-y_2)\quad\mbox{and}\quad \cos{2n\pi\over L}(x_1-x_2)-\cos{2n\pi\over L}(y_1-y_2)\ .$$
We will show that such eigenstates exist for any $N\geq 2$ in any dimension $d\geq 2$
and will find all of them for a system of bosons confined in a $L\times\cdots\times L$
torus $\Lambda$. That is, we will describe all
$N$-particle eigenstates $\psi(\r_1,\ldots,\r_N)$ of the $d\geq 2$-dimensional
kinetic energy operator with the following properties:
(P1) {\it $\psi$ is $L$-periodic separately in each of the $dN$ coordinates}, (P2) {\it $\psi$ is
a symmetric function of $\r_1,\ldots,\r_N$} and (P3)
$\psi(\ldots,\r,\ldots,\r,\ldots)\equiv 0$.
To satisfy (P2) and (P3) at the same time, a very large number of degenerate 
free-particle eigenstates will have to be combined. This becomes possible
because the kinetic energy operator is unbounded and so is 
the degeneracy of the eigensubspaces. On the other hand, this explains why
the energy of these states must increase so fast with $N$.
As the proof of Theorem 2
will show, an interesting interpretation
can be given to these functions: they can be considered 
as stationary states of $N$ impenetrable bosons on the 
$d$-torus which interact with
each other through elastic collisions,
see equations (\ref{pair}) and (\ref{pal}) below.
If $\psi_1,\psi_2,\ldots$ is an orthonormal basis for (P1-3) eigenstates
with corresponding energy eigenvalues $E_1\leq E_2\leq \cdots$, we can define a self-adjoint operator
\be{Helast}
H_{\rm elastic}=\sum_{i=1}^\infty E_i|\psi_i\rangle\langle\psi_i|
\ee
which, hence, is associated to such a system. Also, every member of the family (\ref{H}) can be decomposed
as $H_N=H'_N+H_{\rm elastic}$, with $H'_N$ depending on the interaction and living on the orthogonal
complement of the subspace of (P1-3) eigenstates.

The problem in continuous space, that we have discussed above, will be presented in Section 2. In Section 3
we separately treat the lattice case.
We shall consider only hypercubic lattices.
The analog of (\ref{H}) is obtained by interpreting $\Delta_i$ as the lattice Laplacian.
In lattices contact interactions mean on-site interactions
which exist in any dimension and define the so-called Bose-Hubbard Hamiltonians.
Incidentally, we shall obtain common exact eigenstates in any dimension $d\geq 2$ for a
larger family of Hamiltonians,
\bea{Hlattice}
H_{\rm lattice}=\pm
%\sum_{\r\in \Lambda}\sum_{|\r'-\r|=1}(a^*_{\r'}-a^*_\r)a_\r
\sum_{\langle\r\r'\rangle}(a^*_{\r'}-a^*_\r)(a_{\r'}-a_\r)
+\sum_\r f_\r(n_\r)-\mu\sum_\r n_\r\nonumber\\
+\sum_{i=1}^d\sum_{(\r-\r')\parallel \e_i}f_{\r,\r'}(n_\r,n_{\r'})\ .
\eea
In the first term the summation goes over nearest neighbor pairs,
$a^*_\r$ and $a_\r$ create, resp., annihilate a boson at
site $\r$, $n_\r=a^*_\r a_\r$, $\e_i$ is the unit vector along the $i$th coordinate axis,
$\mu$ is the chemical potential and $f_\r$ and $f_{\r,\r'}$ are real functions with
$f_\r(n)=0$ if $n=0,1$ and $f_{\r,\r'}(m,n)=0$ if $mn=0$.
Among others, every spin-${1\over 2}$ nearest-neighbor XXZ Heisenberg model, including the
planar and the ferro- and antiferromagnetic isotropic models, is a member of this
family (with $f_\r(t)=+\infty$ if $t\geq 2$).
However, there are two limitations in the lattice
case that we do not meet in the continuous one. First, the condition for finding the eigenstates is a
set of transcendental Diophantine equations, in contrast with the algebraic ones in the continuous space
version. We find {\it some} solutions but cannot pretend to find all. Second, 
the solutions we find
are only for $N\leq L^{[d/2]}$ where $[\cdot]$ means integer part,
and this seems to be an intrinsic limitation, due to the boundedness of
$H_{\rm lattice}$. In hard-core models, particle-hole transformation yields eigenstates
also for $N\geq\vol-L^{[d/2]}$.

Finally, we note that both in the continuum and on lattices there is a family of exact eigenstates
whose energy tends to zero as the volume increases. In two and three dimensions
these states may contain $N=o(L^{2/3})$ particles. They correspond to gapless excitations
above the grand canonical ground state if this latter belongs to $N=0$ and, thus, has zero energy.
This is the case e.g. if the interaction is repulsive and the chemical potential is zero or negative
(and we take the plus sign in (\ref{Hlattice})), but may also occur with attractive interactions, as
in the isotropic Heisenberg ferromagnet.

\section{Bosons in continuous space}
\subsection{Theorems and proofs}

For spinless fermions all states share the property (P3). This observation is the starting point of
the construction of such eigenstates for (spinless) bosons in more than one dimension.

We introduce the short-hands $\R=(\r_1,\ldots,\r_N)$ and $\K=(\k_1,\ldots,\k_N)$.
A pair $\K$, $\K'$ will be called {\it allowed} if both sets contain $N$ different vectors
and for any $i$ and $j$, $\k_i+\k'_j$ is in $\Lambda^*\equiv(2\pi/L)\Zz^d$,
where $\Zz^d$ denotes the set of $d$-dimensional integers.
We have the following.

{\bf Lemma 1}
{\it A pair $\K$, $\K'$ is allowed if and only if both sets contain $N$ different vectors, and
there exist $d$-dimensional integers $\n_i$, $\n'_i$ and a (complex) vector $\xi$
such that $\k_i=(2\pi/L)\n_i+\xi$ and $\k'_i=(2\pi/L)\n'_i-\xi$ for every $i$.}

If both $\K$ and $\K'$ contain $N$ different vectors of the above form, the pair is clearly allowed.
On the other hand, if the pair $\K$, $\K'$ is allowed, choose e.g.
$\xi=\k_1$. Then
$\k'_i+\xi=\k_1+\k'_i$ is in
$\Lambda^*$ for every $i$ and $\k_i-\xi=(\k_i+\k'_1)-(\k'_1+\k_1)$ is also in the
Abelian group $\Lambda^*$ for every $i$.

Let us consider the product of two Slater determinants,
\be{Slkk'}
\psi_{\K,\K'}(\R)=\det[e^{i\k_l\r_m}]\det[e^{i\k'_l\r_m}]=\sum_{\pi,\pi'\in S_N}(-)^\pi(-)^{\pi'}
e^{i\sum_{j=1}^N(\k_{\pi(j)}+\k'_{\pi'(j)})\r_j}\ .
\ee
For any allowed pair $\K$, $\K'$, $\psi_{\K,\K'}$ has the properties (P1-3). It is also
an eigenstate of $-\sum\Delta_i$, provided that $\k_i\cdot\k'_j=0$ for any $i,j$, a condition which
can be satisfied above one dimension:
Since $\sum_i(\k_{\pi(i)}+\k'_{\pi'(i)})^2=\sum_i(\k_i+\k'_{\pi'\pi^{-1}(i)})^2$, the $(N!)^2$ plane
waves in (\ref{Slkk'}) may belong to at most $N!$ different eigenvalues. Thus, $\psi_{\K,\K'}$ is an
eigenstate if
$\sum_i(\k_i+\k'_{\pi(i)})^2=\sum_i(\k_i+\k'_i)^2$ or, equivalently,
\be{eigenvalue}
\sum_{i=1}^N\k_i\cdot\k'_{\pi(i)}=\sum_{i=1}^N\k_i\cdot\k'_i
\ee
for all permutations $\pi\in S_N$, which trivially holds if every $\k_i$ is orthogonal to every $\k'_j$.
The following theorem
suggests that there can be a huge redundancy in these equations. It extends the above
example to further eigenstates and shows that for $\K$ and $\K'$ it suffices to satisfy
only $(N-1)^2$ equations.

Below we use the notation $(i_1 i_2 \ldots i_m)$ for the cyclic permutation
carrying $i_1$ into $i_2$, etc., $i_m$ into $i_1$.

{\bf Theorem 1}
{\it Let $\K$, $\K'$ be an allowed pair. The following assertions are
equivalent.\\
(i) Equation (\ref{eigenvalue})
holds for $\pi=(j\,j\+1)$ with $1\leq j\leq N-1$, for
$\pi=(j\,j\+1\,j\+2)$ and
$(j\,j\+2\,j\+1)$ with $1\leq j\leq N-2$ and for 
$\pi=(j\,l\,j\+1\,l\+1)$ and
$(j\,l\+1\,j\+1\,l)$ with $1<j+1<l<N$.\\
(ii)
\be{orthog}
(\k_{j+1}-\k_j)\cdot(\k'_{l+1}-\k'_l)=0\quad,\quad 1\leq j,l\leq N-1\ .
\ee
(iii) Equation (\ref{eigenvalue}) holds true for all permutations.
As a consequence, $\psi_{\K,\K'}$ is an eigenstate of the kinetic energy belonging to
the eigenvalue $\sum(\k_i+\k'_i)^2$ whose degeneracy is at least $(N!)^2$: $\psi_{\K,\K'}$ is the
linear combination of $(N!)^2$ different, pairwise orthogonal, $N$-particle
plane waves.
Furthermore, each occurring plane wave is associated with $N$ different 
wave vectors
chosen from  $\{\k_i+\k'_j\}_{i,j=1}^N$ whose all the $N^2$ elements
are different.}

Since (iii) obviously implies (i), it suffices to show that (ii) follows from
(i) and implies (iii).\\
{\it (i)$\Rightarrow$(ii).}
Observe that we impose (\ref{eigenvalue}) for $(N-1)^2$ different permutations,
and this is just the number of equations (\ref{orthog}) we want to prove.
Writing (\ref{eigenvalue}) for $\pi=(j\,j\+1)$ we immediately obtain
$(\k_{j+1}-\k_j)\cdot(\k'_{j+1}-\k'_j)=0$. If $E_\pi$ denotes the equation
(\ref{eigenvalue}) then $E_{(j\,j\+1\,j\+2)}-E_{(j\,j\+1)}-E_{(j\+1\,j\+2)}$
yields $(\k_{j+2}-\k_{j+1})\cdot(\k'_{j+1}-\k'_j)=0$ and 
$E_{(j\,j\+2\,j\+1)}-E_{(j\,j\+1)}-E_{(j\+1\,j\+2)}$ yields
$(\k_{j+1}-\k_j)\cdot(\k'_{j+2}-\k'_{j+1})=0$. Moreover,
$E_{(j\,l\,j\+1\,l\+1)}-E_{(j\,l)}-E_{(j\+1\,l\+1)}$ is equivalent with
$(\k_{l+1}-\k_l)\cdot(\k'_{j+1}-\k'_j)=0$ from which the last equation
follows by interchanging $j$ and $l$. 

{\it (ii)$\Rightarrow$(iii).}
First, we note that (\ref{orthog}) implies $(\k_j-\k_l)\cdot(\k'_m-\k'_n)=0$
for all $j,l,m,n$. If $\pi=(j\,l)$, (\ref{eigenvalue}) is equivalent with
$(\k_j-\k_l)\cdot(\k'_j-\k'_l)=0$. For a general $\pi$ different from the
identity we write down the equation
(\ref{eigenvalue}) for $\pi$ and for $\pi^{-1}$ (which may be the same)
and rearrange them as
$\sum \k_i\cdot(\k'_{\pi(i)}-\k'_i)=0$ and 
$\sum\k_{\pi(i)}\cdot(\k'_i-\k'_{\pi(i)})=0$,
respectively. The sum of them yields
$\sum(\k_i-\k_{\pi(i)})\cdot(\k'_{\pi(i)}-\k'_i)=0$ which holds because each term
separately vanishes. The difference gives 
$\sum(\k_i+\k_{\pi(i)})\cdot(\k'_{\pi(i)}-\k'_i)=0$ from which we subtract
$2\k_1\cdot\sum(\k'_{\pi(i)}-\k'_i)=0$. Again, the resulting equation holds
because each term equals zero.
It remains to show that in $\psi_{\K,\K'}$ there are $(N!)^2$ different plane waves
and in each of them $N$ different wave vectors.
For this we note that all the $N^2$ wave vectors $\k_i+\k'_j$ are different:
$\k_i+\k'_j=\k_l+\k'_m$ with $(i,j)\neq (l,m)$ would be in conflict with (\ref{orthog}).
Now the coincidence of two plane
waves would mean $\k_{\pi_1(i)}+\k'_{\pi'_1(i)}=\k_{\pi_2(i)}+\k'_{\pi'_2(i)}$
for $\pi_1\neq\pi_2$, $\pi'_1\neq\pi'_2$ and for all $i$, which is impossible;
nor can we have $\k_i+\k'_{\pi(i)}=\k_j+\k'_{\pi(j)}$ unless $i=j$.
This ends the proof.

Let us remark that (\ref{orthog}) is equivalent with
\be{encons}
\epsilon(\k_j+\k'_l)+\epsilon(\k_{j+1}+\k'_{l+1})=\epsilon(\k_j+\k'_{l+1})+\epsilon(\k_{j+1}+\k'_l)
\ee
where $\epsilon(\p)$ is the one-particle energy, $\p^2$. That is, the two-particle energy is
conserved during an elastic
collision of two particles, see also later.

The general solution of (\ref{orthog}) can easily be obtained. The vectors
$\k_{i+1}-\k_i$ and $\k'_{j+1}-\k'_j$ must be in different, orthogonal,
subspaces of $\Rr^d$. One vector from each set, for example $\k_1$ and $\k'_1$, can be
arbitrarily chosen from $\Lambda^*$.
In two dimensions both subspaces are one-dimensional and we have
\be{2Dcont}
\k_i=\k_1+\frac{2\pi}{L}p_i(l_0,m_0)\ ,\quad
\k'_i=\k'_1+\frac{2\pi}{L}p'_i(m_0,-l_0)\quad (i=2,\ldots,N)\ .
\ee
Here $l_0$ and $m_0$ are integers at least one of which is
different from zero, $p_2,\ldots,p_N$ are different nonzero integers and
the same holds for $p'_2,\ldots,p'_N$.
In three dimensions one of the subspaces is one-, the other is two-dimensional.
Without restricting generality we may suppose that $\k_j-\k_i$ are in a
one-dimensional subspace. Let $\n_0$ be any nonzero integer vector. Since the
subspace orthogonal to $\n_0$ has an infinite intersection with
$\Zz^3$, there is no problem to choose the vectors of $\K'$. So we have
\be{3Dcont}
\k_i=\k_1+{2\pi\over L}p_i\n_0\ ,\quad
\k'_i=\k'_1+{2\pi\over L}\n'_i\quad
(i=2,\ldots,N)\ ,
\ee
where $p_i$ are different nonzero integers
and $\n'_i$ are different nonzero integer vectors orthogonal to $\n_0$.

A complex vector $\xi$ (the same one which has appeared in Lemma 1)
can be added to $\k_1$ and subtracted from $\k'_1$ without changing
the eigenvalue or the eigenfunction: it is a kind of gauge variable which affects only
the decomposition of $\psi_{\K,\K'}$ in the product of two determinants.

We note that the two examples given in the Introduction correspond to $\psi_{\K,\K'}$
with $\K=\pm{2\pi\over L}(n,0)$, $\K'=\pm{2\pi\over L}(0,m)$ and
$\K=\pm{\pi\over L}(n,n)$, $\K'=\pm{\pi\over L}(n,-n)$, respectively.

The question naturally arises whether with Theorem 1 we have described all the
eigenstates of the kinetic energy operator satisfying (P1-3). For the moment we cannot even exclude
the existence of other eigenstates
$\psi_{\K,\K'}$ which do not fall under the authority of Theorem 1.
For a general $\psi_{\K,\K'}$ with an allowed pair $\K,\K'$, some elements of $\{\k_i+\k'_j\}_{i,j}$ may
coincide and,
thus, the number of different wave vectors can be less than $N^2$. This may also imply
that, by cancellation, the number of different $N$-particle plane waves contributing to $\psi_{\K,\K'}$
is less than $(N!)^2$. (Cancellation occurs if
$\{\k_i+\k'_{\alpha(i)}\}_{i=1}^N=\{\k_i+\k'_{\beta(i)}\}_{i=1}^N$ for two permutations $\alpha$ and $\beta$
of different parity. It is easily verified that for $N=2$ and 3 this is not possible but can take place for
$N\geq 4$. An example for $d=2$ and $N=4$ is $\K={2\pi\over L}((0,-1),(1,0),(0,1),(-1,0))$ and
$\K'={2\pi\over L}((-2,1),(1,1),(-1,0),(0,0))$ with $\alpha=(2\,4)$ and $\beta=(1\,2)(3\,4)$.)
As a consequence, some of the equations (\ref{eigenvalue}) could not be satisfied but $\psi_{\K,\K'}$
still could be an eigenstate. In the following, we exclude this and all other possibilities.

{\bf Theorem 2}
{\it For $d>1$, among
the eigenstates of the kinetic energy which satisfy the conditions
{\rm (P1-3)} the set of those described in Theorem 1 contains a basis.
In one dimension such eigenstates do not exist.}

To begin the proof, we note that
an eigenstate of the kinetic energy satisfying (P1) and (P2) is a finite
linear combination of symmetrized plane waves,
\be{N}
\psi(\R)=\sum_{\alpha\in I} a_{\alpha}\psi^+_{\Pal}(\R)\ .
\ee
In this expression $\alpha$ goes over some index set $I$, $\Pal=(\pal_1,\ldots,\pal_N)$,
\be{psi+}
\psi^+_{\Pal}(\R)=\sum_{\pi\in S_N} e^{i\sum_j\p^\alpha_{\pi(j)}\r_j}
\ee
and, for each $\alpha$, all $\p^\alpha_i$ are in $\Lambda^*$ and
$\sum_i(\p^\alpha_i)^2$ is independent of $\alpha$. 
If $a_\alpha$ are complex, $\psi=\psi_{\rm Re}+i\psi_{\rm Im}$ where
$\psi_{\rm Re}=\sum_{\alpha\in I}({\rm Re\,}a_\alpha)\psi^+_\Pal$ and
$\psi_{\rm Im}=\sum_{\alpha\in I}({\rm Im\,}a_\alpha)\psi^+_\Pal$. It is
easily seen that (P3) holds for $\psi$ if and only if it holds both for
$\psi_{\rm Re}$ and $\psi_{\rm Im}$. Therefore, when looking for a basis among (P1-3) eigenstates, we can limit
our considerations to functions (\ref{N}) with real coefficients. To further specialize our choice of a basis,
we need the notion of minimality.
When writing (\ref{N}), we understand that all the possible contractions in $\psi$
have been made, so that $a_\alpha\neq 0$ if $\alpha\in I$ and the sets
$\{\pal_i\}_{i=1}^N$ are different for different $\alpha$ (but may overlap with each other).
We call the orthogonal set $\{\psi^+_\Pal\}_{\alpha\in I}$ of (P1-2)
eigenstates, all belonging to the same energy, the {\it support}
of $\psi$ and denote it by supp$\,\psi$. A (P1-3) eigenstate will be called a
{\it minimal function} if it is the unique (P1-3) eigenfunction
(apart from a constant multiplier) in the linear span of its own support.

It will suffice to study only minimal functions because of the following.

{\bf Lemma 2}
{\it Among the (P1-3) eigenstates there exists a basis formed by minimal functions.}

Suppose that $\phi$ is a non-minimal (P1-3)-eigenfunction. It is enough to show that it can be
decomposed as $\phi=a\phi_1+b\phi_2$ where $\phi_1$ and $\phi_2$ are (P1-3)-eigenfunctions whose
support is strictly smaller than that of $\phi$. Then, if $\phi_i$ is not minimal, it can be decomposed
in the same manner as $\phi$ and, in a finite number of steps, we arrive at an expression of
$\phi$ as a sum of minimal functions. Now,
because $\phi=\sum_{\alpha\in I}a_\alpha\psi^+_\Pal$ is not minimal, its support spans an at least
two-dimensional subspace of (P3)-eigenstates. Take here a
$\phi'=\sum_{\alpha\in I}b_\alpha\psi^+_\Pal$ which is linearly independent
of $\phi$ (and may have vanishing coefficients). Choose a $\beta$ such that 
$b_\beta\neq 0$ and define $\phi_1=\phi-{a_\beta\over b_\beta}\phi'$. Take a 
$\gamma\neq\beta$ and define $\phi_2=\phi'-{b_\gamma\over a_\gamma}\phi$.
Then 
\be{min}
\phi=\left(1-{a_\beta b_\gamma\over a_\gamma b_\beta}\right)^{-1}\phi_1
+\left({b_\beta\over a_\beta}-{b_\gamma\over a_\gamma}\right)^{-1}\phi_2
\ee
is the decomposition we were looking for.
                                                                                    
From now on, we shall suppose that $\psi$ also satisfies (P3) and is minimal.
We are going to prove that {\it the minimal functions
are precisely the eigenstates described in Theorem 1}.

First we consider $N=2$. The simplest (P1-3)-eigenstates are of the form
\bea{N=2}
\psi(\r_1,\r_2)&=&e^{i(\p_1\r_1+\p_2\r_2)}+e^{i(\p_2\r_1+\p_1\r_2)}
-e^{i(\q_1\r_1+\q_2\r_2)}-e^{i(\q_2\r_1+\q_1\r_2)}\nonumber\\
               &=&\psi^+_\P(\r_1,\r_2)-\psi^+_\Q(\r_1,\r_2)
\eea
where $\p_1,\p_2,\q_1,\q_2\in \Lambda^*$ and
\be{pair}
\p_1+\p_2=\q_1+\q_2\quad,\quad \p_1^2+\p_2^2=\q_1^2+\q_2^2
\ee
or equivalently,
\be{pair'}
\p_1+\p_2=\q_1+\q_2\quad,\quad \p_1\cdot\p_2=\q_1\cdot\q_2\ .
\ee
These eigenstates are certainly minimal, because a (P3)-eigenstate must contain at least two
symmetrized plane waves.

In any dimension equations (\ref{pair}) have trivial solutions,
$\p_1=\q_1$ and $\p_2=\q_2$ with $\p_i\in\Lambda^*$ arbitrary,
which are useless because they yield $\psi(\r_1,\r_2)\equiv 0$.
Thus, we must impose
$\{\p_1,\p_2\}\cap\{\q_1,\q_2\}=\emptyset$, since an overlap
would imply coincidence of the two sets. Moreover, if $\p_1=\p_2=\p$
then $\q_1=\q_2=\p$, a solution already discarded.
Hence, in any nontrivial solution of (\ref{pair}), $\p_1,\p_2,\q_1$ and $\q_2$ are
four different vectors. 

In one dimension, if any of the four numbers is zero,
we obtain a trivial solution. In the opposite case $q_2=p_1p_2/q_1$. This
we insert in the first of Eqs. (\ref{pair}) to find $(p_1-q_1)(q_1-p_2)=0$
which, again, leads to a trivial solution.

For $d>1$ let $\p_1$, $\p_2$, $\q_1$ and $\q_2$ be any nontrivial solution
of Eqs. (\ref{pair'}).
Choose an arbitrary vector $\k'_2$ and introduce
$\k_1=\q_1-\k'_2$, $\k_2=\p_2-\k'_2$ and $\k'_1=\p_1-\q_1+\k'_2$.  
Then $\k_1\neq\k_2$ and $\k'_1\neq\k'_2$. The original vectors can be written in the form
$\p_1=\k_1+\k'_1$, $\p_2=\k_2+\k'_2$, $\q_1=\k_1+\k'_2$ and, because of the
first of Eqs. (\ref{pair'}), $\q_2=\k_2+\k'_1$. Furthermore, (\ref{pair'}) implies
$(\k_2-\k_1)\cdot(\k'_2-\k'_1)=(\p_2-\q_1)\cdot(\q_1-\p_1)=0$. Thus, the eigenstate (\ref{N=2})
takes on the form prescribed in Theorem 1. Clearly,
$\k'_2$ plays the role of the gauge vector $\xi$.

It remains to show that minimal eigenfunctions, being the sum of more than
two symmetrized plane waves, do not exist for $N=2$. Let us suppose
the opposite, i.e. that for some $n>2$ there is a
minimal eigenfunction $\psi(\r_1,\r_2)=\sum_{m=1}^n a_m \psi^+_{\P^m}$,
where $\P^m=(\p^m_1,\p^m_2)$
and $(\p^m_1)^2+(\p^m_2)^2$ is independent of $m$.
Because of the minimality, identical vanishing of
$\psi(\r,\r)=2\sum_{m=1}^n a_m e^{i(\p^m_1+\p^m_2)\r}$ implies that also
$\p^m_1+\p^m_2$ is independent of $m$ (indeed
possible for $n>2$, see Lemma 3 below), and $\sum_{m=1}^n a_m=0$. But then
$\psi(\r_1,\r_2)=\sum_{m=1}^{n-1}a_m[\psi^+_{\P^m}(\r_1,\r_2)
-\psi^+_{\P^n}(\r_1,\r_2)]$, where each term of the sum is a minimal eigenfunction
of the type (\ref{N=2}),
so $\psi$ could not be minimal. Thus we have shown the theorem for $N=2$.

We complete the study of the case $N=2$ by answering the following question: Given $\p_1$ and
$\p_2\neq \p_1$, both in $\Lambda^*$, how to obtain all the solutions of Eqs. (\ref{pair}) for $\q_1$ and $\q_2$?

{\bf Lemma 3}
{\it Let $\p_1\neq\p_2$, both in $\Lambda^*$, and consider the sphere $S$ in $\Rr^d$ drawn over
$\p_1-\p_2$ as a diameter. There is a one-to-one correspondence between nonzero orthogonal pairs
of vectors in the intersection of $S$
with $\Lambda^*$ and the solutions of equation (\ref{pair}) for $\{\q_1,\q_2\}$. If
$\a$ and $\b$ form such a pair, the corresponding solution is
$\q_1=\p_1-\a=\p_2+\b$ and $\q_2=\p_2+\a=\p_1-\b$.}

If there is a pair $\{\q_1,\q_2\}\neq\{\p_1,\p_2\}$ which solves (\ref{pair'}), we use the
parametrization with an allowed pair $\K,\K'$ described above
and define $\a=\k'_1-\k'_2$ and $\b=\k_1-\k_2$. Then $\q_1$ and $\q_2$ are as asserted,
$\a\cdot\b=0$ and $\p_1-\p_2=\a+\b$.
As we have seen in Lemma 1, both $\a$ and $\b$ are in $\Lambda^*$ and, because they are
orthogonal and sum up to $\p_1-\p_2$, they are also on the sphere
\be{sphere}
S=\{\r\in\Rr^d:|\r-\frac{1}{2}(\p_1-\p_2)|=\frac{1}{2}|\p_1-\p_2|\}\ .
\ee
Oppositely,
let $\a\in\Lambda^*\cap S$ and define $\q_1=\p_1-\a$ and $\q_2=\p_2+\a$. Then $\q_1+\q_2=\p_1+\p_2$
and $\q_1\cdot\q_2=\p_1\cdot\p_2+\a\cdot(\p_1-\p_2)-\a^2=\p_1\cdot\p_2$ because $\a\in S$. Now
along with $\a$, $\b=\p_1-\p_2-\a$ is also in the intersection and
$\a\cdot\b=0$. So vectors in $\Lambda^*\cap S$ occur in orthogonal pairs whose elements add up to $\p_1-\p_2$
and provide the same solution.

Any nontrivial partition of $(2\pi/L)^{-2}|\p_1-\p_2|^2$ into the sum of the square of
$2d$ integers gives rise to at least one solution of (\ref{pair}). 
Therefore, if $\p_1-\p_2$ has two nonvanishing coordinates, there exists at least one orthogonal
pair of vectors described in the lemma. On the other hand, if only a single coordinate is nonvanishing,
there may not be any nontrivial solution for $\q_1,\q_2$ (take, e.g. $d=2$
and $\p_1-\p_2=(2\pi/L)(0,3)$).

We turn to the case $N>2$.
Since $\psi(\R)$ is a symmetric function, for (P3) to hold $\psi(\r,\r,\r_3,\ldots)\equiv 0$ suffices.
To see how $\psi(\r,\r,\r_3,\ldots)$ can vanish, we write
\be{psi+2}
\psi^+_{\Pal}(\R)=\sum_{1\leq l<m\leq N}\psi^{lm}_\Pal(\R)
\ee
with
\bea{psi+3}
\psi^{lm}_\P(\R)&=&\left[e^{i(\p_l\r_1+\p_m\r_2)}+e^{i(\p_m\r_1+\p_l\r_2)}
\right]\sum_{\pi\in S_N:\pi(1)=l,\pi(2)=m}e^{i\sum_{n=3}^N\p_{\pi(n)}\r_n}
\nonumber\\
&=&\psi^+_{\{\p_l,\p_m\}}(\r_1,\r_2)\psi^+_{\{\p_j\}_{j\neq l,m}}(\r_3,
\ldots,\r_N)\ .
\eea
Here we have used again the notation (\ref{psi+}) for symmetrized plane waves.
Now $\psi^{lm}_\Pal(\r,\r,\r_3,\ldots)$ can be cancelled as a whole by
an identically equal $\psi^{st}_\Pbet(\r,\r,\r_3,\ldots)$.
The condition of an identical equality imposes
\be{palset}
\{\pal_j\}_{j\neq l,m}=\{\pbet_j\}_{j\neq s,t}
\ee
and
\be{pal}
\pal_l+\pal_m=\pbet_{s}+\pbet_{t}\quad,\quad
(\pal_l)^2\+(\pal_m)^2=(\pbet_{s})^2\+(\pbet_{t})^2
\ee
where we have added the last equality which comes from the fact that 
$\psi^+_\Pbet$ must belong
to the same energy eigenvalue as $\psi^+_\Pal$. So for the vectors
$\pal_l,\pal_m,\pbet_{s},\pbet_{t}$ we find precisely the equations
(\ref{pair}). Again, we have to discard the trivial solutions. They would yield
$\Pal=\Pbet$, so
$\psi^+_\Pal(\R)\equiv\psi^+_\Pbet(\R)$,
which contradicts the supposed contracted form of $\psi$.
Considering equation (\ref{pal}) for every pair $l<m$, we conclude that for each $\alpha$ the
set $\{\pal_j\}_{j=1}^N$ contains $N$ different vectors. Furthermore,
if $\psi_\Pal^{l_1m_1}(\r,\r,\r_3,\ldots)\equiv\psi_\Pbet^{s_1t_1}(\r,\r,\r_3,\ldots)$ then
$\{\pal_j\}_{j=1}^N\cap\{\pbet_j\}_{j=1}^N=\{\pal_j\}_{j\neq l_1,m_1}
=\{\pbet_j\}_{j\neq s_1,t_1}$,
and for the same $\alpha$ and $\beta$ no other choice of $l<m$ and $s<t$ can 
yield the equality (\ref{palset}).
Thus, to annihilate $\psi^+_\Pal(\r,\r,\r_3,\ldots)$ we need one term
from the decomposition (\ref{psi+2}) of
$N(N-1)/2$ different $\psi^+_\Pbet$. Solving the $N(N-1)/2$ equations (\ref{pal}) for $\{\pbet_s,\pbet_t\}$,
one for each pair
$l<m$, we may (but for the moment we cannot say we really do)
obtain $N(N-1)$ different vectors which, together with the $N$ vectors of $\Pal$, form a set of
$N^2$ wave vectors. Recall that $\psi_{\K,\K'}$ of Theorem 1 are also constructed with the help of
$N^2$ vectors $\k_i+\k'_j$. So far we have found that minimal eigenfunctions share some attributes of
$\psi_{\K,\K'}$, but have not seen as yet that these latter themselves are minimal.

{\bf Lemma 4}
{\it $\psi_{\K,\K'}$ of Theorem 1 are minimal (P1-3)-eigenfunctions.}

In the present case the index set is $S_N$.
Let $(\K+\alpha\K')_j=\k_j+\k'_{\alpha^{-1}(j)}$.
We have to prove that
\be{psikk'}
\psi_{\K,\K'}=\sum_{\alpha\in S_N}(-)^\alpha\psi^+_{\K+\alpha\K'}
\ee
is the unique (P1-3)-eigenstate that can be obtained as a
linear combination of the functions $\psi^+_{\K+\alpha\K'}$ with $\alpha$
going over $S_N$. According to equations (\ref{N}), (\ref{psi+2}) and
(\ref{psi+3}), a general function over the above basis reads
\bea{Sl2}
\psi(\R)=\sum_{\alpha\in S_N}a_{\alpha}\psi^+_{\K+\alpha\K'}(\R)
=\sum_{\alpha\;{\rm even}}\sum_{l<m}\left[a_\alpha\psi^+_{\{(\K+\alpha\K')_j\}_{j=l,m}}(\r_1,\r_2)\right.\nonumber\\
\left.+a_{(lm)\alpha}\psi^+_{\{(\K+(lm)\alpha\K')_j\}_{j=l,m}}(\r_1,\r_2)\right]
\psi^+_{\{(\K+\alpha\K')_j\}_{j\neq l,m}}(\r_3,\ldots)\ .
\eea
We may suppose that the coefficient of the identity is nonzero -- this can always be achieved by
renumbering the vectors of $\K'$, if necessary -- and
normalize $\psi$ so as to have $a_{\rm id.}=1$.
In (\ref{Sl2}) we have regrouped the terms which can cancel each other if $\r_1=\r_2$.
The point is that, given
$(\alpha,l<m)$, there is a unique $(\beta,s<t)$ such that equations (\ref{palset}) and (\ref{pal}) hold
true, namely, $\beta=(l\,m)\alpha$, $s=l$ and $t=m$. This is because $\k_i+\k'_j$ uniquely determines
$(i,j)$. Thus, if $\psi$
is a (P3)-eigenstate then for every $\alpha$ and every pair $l<m$, $a_{(lm)\alpha}=-a_\alpha$.
Starting with $(l\,m)\in S_N$, $a_{(lm)}=-1$, and the products of any two, three, etc. of them appear
with coefficients $(-)^2$, $(-)^3$, etc., respectively. Since
the inversions generate $S_N$, we conclude that $a_\alpha=(-)^\alpha$ for each $\alpha\in S_N$
and, thus, $\psi=\psi_{\K,\K'}$, proving that the latter is minimal.

The last step is to show that no other type of minimal eigenfunctions exists. 
This we have seen for
two particles and will prove for $N\geq 3$ by induction. Suppose the claim is verified for any
$n<N$. Let $\calp=\cup_{\alpha\in I}\{\pal_i\}_{i=1}^N$ i.e. the full set of wave vectors
used to define the minimal (P1-3) eigenfunction (\ref{N}). After resummation this function reads
\be{Nbis}
\psi(\R)=\sum_{\p\in\calp}b_\p e^{i\p\cdot\r_N}\phi_\p(\r_1,\ldots,\r_{N-1})
\ .
\ee
Here every $\phi_\p$ is a uniquely determined (P1-3) eigenfunction of 
$-\sum_{i=1}^{N-1}\Delta_i$ which,
by the induction
hypothesis, is a linear combination of $N-1$-variable eigenfunctions of the 
type described in Theorem 1.
(That $\phi_\p$ is actually minimal and, thus,
equals a single function of this type, will follow from the proof.)
The sum (\ref{Nbis}) can be considered as an embedding of any given of its 
terms into a minimal (P1-3) eigenfunction.
The $(N-1)^2$ wave vectors appearing in each minimal component of $\phi_\p$
are of the form
$\k_i+\k'_j$ with $1\leq i,j\leq N-1$ and satisfy
the orthogonality relations (\ref{orthog}). 
Consider the function
$$\sum_{\pi,\pi'\in S_{N-1}}(-)^{\pi\pi'}\exp\{i\p\r_N
+i\sum_{j=1}^{N-1}(\k_{\pi(j)}+\k'_{\pi'(j)})\r_j\}$$
in the term $\p$ of (\ref{Nbis}).
The symmetrical embedding of the plane wave belonging to 
$(\pi,\pi')$ implies the presence, in the expansion of $\psi$, 
$N-1$ other plane waves, one for each $j$, in which
$\p$ and $\k_{\pi(j)}+\k'_{\pi'(j)}$ are interchanged.
So $\k_{\pi(j)}+\k'_{\pi'(j)}\in\calp$ for $j=1,\ldots,N-1$ and
embedding of the $(N-1)$-particle plane wave
$\exp\{i\p\r_j+\sum_{l\neq j}^{N-1}(\k_{\pi(l)}+\k'_{\pi'(l)})\r_l\}$ in 
$\phi_{\k_{\pi(j)}+\k'_{\pi'(j)}}(\r_1,\ldots,\r_{N-1})$
implies that $\p=\k+\k'$, and
$\{(\k,\k')\}\cup\{(\k_j,\k'_j)\}_{j=1}^{N-1}$ satisfy (\ref{orthog}).
We conclude that in $\calp$ there is a subset
parametrized by an allowed 
pair $\K,\K'$, both members containing $N$ vectors such that
(\ref{orthog}) holds, and all the corresponding symmetrized plane waves 
$\psi^+_{\K+\alpha\K'}$
occur in the support of $\psi$. Since $\psi$ is minimal, this {\it is} 
the support of $\psi$, and 
$\psi=\psi_{\K,\K'}$. This ends the proof of the theorem.

\subsection{Remarks}

{\bf 1.}
The representation of a bosonic state as
the product of two Slater determinants is, in a certain sense, natural.
Intuitively, we can
imagine a spinless boson to be built up of two coinciding half-spin fermions of
opposite spins, and the
two Slater determinants as wavefunctions of the two kinds of fermions.
Also,
the eigenstates we are looking for are twice continuously differentiable ($C^2$, in fact, analytic)
functions of all
coordinates and vanish, therefore, at least quadratically at $\r_i=\r_j$. The $C^2$-functions of all
coordinates satisfying (P1-3) form a linear space $\mathcal L$, and one may suspect
that $\psi_{\K,\K'}$ with allowed pairs is a generating system of this space. A way to prove it
would be to take an arbitrary function of $\mathcal L$, divide and re-multiply
it with a `nice' antisymmetric function and expand both -- now antisymmetric -- factors as a linear
combination of Slater determinants which are known to form a basis in the antisymmetric subspace.
The crux is the choice of the auxiliary antisymmetric function. It should yield an at least
$L^2$-convergent expansion for both factors, and the product of the two expansions should still be
$L^2$-convergent. Because $\mathcal L$ is dense in the symmetric subspace of $L^2(\Lambda^N)$,
a successful proof would imply that $\psi_{\K,\K'}$ with allowed pairs linearly span the symmetric
subspace.
%Note, however, that this would be a very bad basis for proving off-diagonal long-range order,
%see remark 9 below.

{\bf 2.}
The form of $\psi_{\K,\K'}$, corresponding to (\ref{Nbis}), is
\be{embed}
\psi_{\K,\K'}(\R)=\sum_{l,m=1}^N(-)^{2N-l-m}e^{i(\k_l+\k'_m)\cdot\r_N}
\phi_{\k_l+\k'_m}(\r_1,...,\r_{N-1})\ ,
\ee
where $\phi_{\k_l+\k'_m}=\psi_{\K\setminus\{\k_l\},\K'\setminus\{\k'_m\}}$.

{\bf 3.}
Theorem 2 is not claiming that eigenstates, being, for instance,
the prod\-uct of more than two Slater determinants,
do not exist. It is easily seen that in $d\geq n$ dimensions,
with $n$ pairwise orthogonal k-sets one can obtain eigenstates in the form of
a product of $n$ Slater determinants. However, for $n$ even these states can be expanded in the basis
$\{\psi_{\K,\K'}\}$ (as for $n$ odd they can be expanded in the basis of Slater determinants).
As an example, the expansion of a product of four determinants belonging to
$\K,\K^1,\K^2$ and $\K^3$, respectively, reads
\be{4det}
\psi_{\K,\K^1,\K^2,\K^3}=\sum_{\alpha,\beta\in S_N}(-)^{\alpha\beta}\psi_{\K,\K^1+\alpha\K^2+\beta\K^3}\ ,
\ee
irrespective of the space dimension or whether the product is an
eigenfunction.

{\bf 4.}
The energy of the eigenstates $\psi_{\K,\K'}$ is easy to estimate.
In two and three dimensions the length
of the vectors in one of the two sets has to increase at least linearly, so the energy is at least of the
order of $N^3/L^2$ (because of the orthogonality (\ref{orthog}), at the same time
$$\sum\k_i\cdot\k'_i=\k_1\cdot\sum(\k'_i-\k'_1)
+\k'_1\cdot\sum(\k_i-\k_1)+N\k_1\cdot\k'_1$$
is only of order $N^2/L^2$).
At fixed positive densities this is much larger than the ground state energy 
of any physically meaningful $H_N$ (which is smaller than a multiple of $N$),
and these states play no role in
thermodynamics either. However, in cases when
the grand-canonical ground state is the
vacuum, $N=0$, as e.g.
for repulsive pair interactions and zero or
negative chemical potentials, we obtain gapless excitations in the form of
$N=o(L^{2/3})$-particle eigenstates
whose energy vanishes with the increasing volume.

{\bf 5.}
Lemma 3 shows that degeneracy of minimal states can occur. Because of nonvanishing overlaps,
within a degenerate subspace the
minimal states may not be linearly independent
and one may not choose an orthogonal basis of them.

{\bf 6.}
$\psi_{\K,\K'}$ are eigenstates of the total momentum operator with eigenvalue $N\q=\sum(\k_i+\k'_i)$.
Multiplying and dividing $\psi_{\K,\K'}$ by $\exp i\,\q\sum\r_j$ yields
\be{mom}
\psi_{\K,\K'}(\R)=e^{i\,\q\sum\r_j}\psi_{\K-\kbar,\K'-\kprimbar}(\R)
\ee
with $\kbar=N^{-1}\sum\k_i$ and $\K-\kbar=(\k_j-\kbar)_{j=1}^N$ and similar for the primed variables.
If $\psi_{\K,\K'}$ is an eigenstate of the kinetic energy then $\psi_{\K-\kbar,\K'-\kprimbar}$ is a
zero-momentum eigenstate whose k-sets satisfy not only
(\ref{orthog}) but also the more specific equations $(\k_i-\kbar)\cdot(\k'_j-\kprimbar)=0$ which are easy
to verify. Replacing $\k_i-\kbar$ by $\k_i$ and $\k'_i-\kprimbar$ by $\k'_i$ we obtain the canonical
form of minimal functions:

{\bf Corollary}
{\it A basis of (P1-3) eigenstates can be chosen among the functions
\be{canon}
\psi_{\q,\K,\K'}(\R)=e^{i\q\sum\r_j}\psi_{\K,\K'}(\R)
\ee
where $\q\in\Lambda^*$,
$N\q$ is the total momentum and $\psi_{\K,\K'}$ is a zero-momentum minimal eigenfunction. Namely,
$\K,\K'$ is an allowed pair with $\k_i\cdot\k'_j=0$ for any $i$ and $j$ and $\sum\k_i=\sum\k'_i=0$.}

In the right-hand side of equation (\ref{canon}) $\R$ can be replaced by $(\r_i-\overline{\r})_{i=1}^N$
where $\overline{\r}=N^{-1}\sum_{i=1}^N\r_i$.

{\bf 7.}
Equations (\ref{pair}) and (\ref{pal}) suggest to interprete
the (P1-3) eigenstates of the kinetic energy operator as  stationary states of
a system of impenetrable, pointlike bosons which interact with each other through elastic collisions.
Clearly, there exists no self-adjoint Hamiltonian with a dense domain in the symmetric
subspace of $L^2(\Lambda^N)$ which would describe such a system! The states we have found
cover only a very small fraction (roughly $\sim (N!)^{-2}$) of the Hilbert space. Since
the suitable operator would be $H_N$ with infinite repulsive delta-interactions, Theorem 2 provides
an independent proof of the otherwise known fact, cf. the Introduction,
that above one dimension $-\sum\Delta_i$, defined as a symmetric operator $H^0$ on
$C^\infty_0(\Lambda^N\setminus\cup_{i\neq j}\{\r_i=\r_j\})$, has no self-adjoint extension
describing such an interaction.
The states characterized in the above theorems are in the domain of $\overline{H^0}$, the closure of
$H^0$.
In two and three dimensions they are
eigenstates of any Hamiltonian $H_N$
with a properly defined contact interaction, and the only eigenstates in the domain of
$\overline{H^0}$.
Any such $H_N$ is a self-adjoint extension of $H^0$ or
$\overline{H^0}$, so
densely defined in $L^2(\Lambda^N)$, and having infinitely many other symmetric
eigenstates, which do {\it not}
vanish (even diverge) at particle encounters. As suggested in the Introduction,
we can orthogonally decompose any $H_N$ in an interaction-dependent part and $H_{\rm elastic}$,
cf. equation (\ref{Helast}).

{\bf 8.}
Because of their simple form, some basic properties of minimal functions can be studied easily.
Nodal surfaces, for example, can be identified if the function
is written in the canonical
form (\ref{canon}). Let $V$ and $V'$ be the orthogonal subspaces containing 
$\k_j$ and $\k'_j$, respectively.
Since $\det[e^{i\k_l\r_m}]\equiv 0$ if $\r_i-\r_j$ is in $V'$ and $\det[e^{i\k'_l\r_m}]\equiv 0$ if
$\r_i-\r_j$ is in $V$, both for any $i\neq j$,
we find that $\psi_{\K,\K'}(\R)\equiv 0$ if $\r_i-\r_j$ is in $V\cup V'$ 
for a pair
$i\neq j$. So there is a regular long-range exclusion effect, although it concerns only a set of zero
Lebesgue measure in the configuration space. This effect can be amplified by
particular choices of
$\K$ or $\K'$, that we can see easier on the two-point correlation function. 
A straightforward computation of this latter yields
\bea{corr}
\rho_2(0,\r)
\equiv N(N-1)\langle\delta(\r_1)\delta(\r_2-\r)\rangle\,
\phantom{aaaaaaaaaaaaaaaaaaaaa}\nonumber\\
=\rho^2{N\over N-1}\left(1-{1\over N^2}\left|\sum_le^{i\k_l\r}\right|^2\right)
\left(1-{1\over N^2}\left|\sum_le^{i\k'_l\r}\right|^2\right)
\eea
where $\rho=N/\vol$ is the uniform value of $\rho_1(\r)=N\langle\delta(\r_1-\r)\rangle$
and averaging means integration over $\r_1,\ldots,\r_N$ in $\Lambda$ with the weight function
$|\psi_{\K,\K'}|^2/\|\psi_{\K,\K'}\|^2$. In the derivation of (\ref{corr}) the minimality of $\psi_{\K,\K'}$
is exploited via the property that $\k_i+\k'_j=\k_l+\k'_m$ implies $i=l$ and $j=m$.
For most choices of $\K$ and $\K'$,
$\rho_2(0,\r)\approx\rho^2$ outside a small neighborhood
of $V\cup V'$ where it vanishes. In general,
let $\calg_\K$ be the set of points $\r$ in $\Lambda$ such that $\k_l\cdot\r$ is an integer multiple of
$2\pi$ for every $l$. $\calg_\K$ is an Abelian group with respect to addition modulo $\Lambda$
which contains $V'$ as a subgroup and may contain translates of $V'$. For example,
for $i=1,\ldots,d$ let
$M_i$ be the greatest common divisor of
$\{(L/2\pi)k_{ni}\}_{n=1}^N$% and $\{(L/2\pi)k'_{ni}\}_{n=1}^N$, respectively,
where $k_{ni}$ is the $i\,$th component of $\k_n$.
%Set $M_i=\infty$ if $k_{ni}=0$ for all $n$, similarly for $M'_i$.
Then $\calg_\K$ contains the points with coordinates
$x_i=m_iL/M_i$, where $m_i$ is any integer between $-M_i/2$ and $M_i/2$.
Defining $\calg_{\K'}$ analogously,
we get another Abelian group containing $V$ and maybe some translates of it.
%and define similarly $\calg'$ with $M'_i$. For $\r$ in $\calg$,
Now $\rho_2(0,\r)=0$ if and only if $\r$ is in $\calg_\K\cup\calg_{\K'}$ and,
as we can check also directly, $\psi_{\K,\K'}(\R)=0$ if (and only if) $\r_i-\r_j$ is in
$\calg_\K\cup\calg_{\K'}$ for a pair $i\neq j$.

{\bf 9.}
In the eigenstates $\psi_{\K,\K'}$ there is no off-diagonal long-range order:
The associated one-particle reduced density matrix, $\sigma$, has $N^2$ nonvanishing eigenvalues,
each of which is equal to $1/N$. Indeed, using that all $\k_l+\k'_m$ are different,
a straightforward calculation of the integral kernel of $\sigma$ yields
\bea{sigma}
\sigma(\r,\r')&=&N\|\psi_{\K,\K'}\|^{-2}\int_{\Lambda^{N-1}}\psi_{\K,\K'}(\r,\r_2,\ldots)
\psi^*_{\K,\K'}(\r',\r_2,\ldots){\rm d}\r_2\ldots{\rm d}\r_N\nonumber\\
            &=&{1\over N\vol}\sum_{l,m=1}^N e^{i(\k_l+\k'_m)(\r-\r')}\ .
\eea
Because $\psi_{\K,\K'}$ is a momentum eigenstate as well, $\sigma$ is diagonal in momentum
representation and its eigenvalues are the diagonal elements. Fourier transforming $\sigma(\r,\r')$,
\be{sigmapp}
\sigma(\p,\p)=\frac{1}{N}\sum_{l,m=1}^N\delta_{\p,\k_l+\k'_m}\ .
\ee
Since $\p=\k_l+\k'_m$ can hold for at most a single pair $(l,m)$, the result
follows.

{\bf 10.}
The theorems extend to systems in $L_1\times\cdots\times L_d$ rectangles with periodic
boundary conditions. The appropriate definition of $\Lambda^*$ is
$(2\pi/L_1)\Zz\times\cdots\times(2\pi/L_d)\Zz$. The orthogonality relation (\ref{orthog}) holds
for example
if $\{\k_i-\k_j\}$ and $\{\k'_i-\k'_j\}$ have disjoint subsets of nonvanishing coordinates.
The existence of other solutions depends on the rationality of the ratios among $L_i$.

\section{Eigenstates for lattice models}

We consider the same problem as before on an $L\times\cdots\times L$ part
(now $L$ is a positive integer) of $\Zz^d$ with periodic boundary conditions.
Defining the lattice Laplacian as
\be{latLap}
(\Delta\psi)(\r)=\sum_{|\r'-\r|=1}[\psi(\r')-\psi(\r)]\ ,
\ee
$e^{i\k\r}$ with $\k=(k_1,\ldots,k_d)$ is an eigenfunction of $-\Delta$ 
belonging to the eigenvalue $\epsilon(\k)=2\sum_{i=1}^d(1-\cos k_i)$.
A pair $\K,\K'$ is called allowed if both
$\{\k_j\}_{j=1}^N$ and $\{\k'_j\}_{j=1}^N$ contain $N$ incongruent
vectors modulo $2\pi$ and $\k_i+\k'_j$ is in $\Lambda^*$ for all $i,j$.
Now for an allowed pair $\K,\K'$, $\psi_{\K,\K'}$ as given by (\ref{Slkk'})
is an eigenstate of $-\sum_{i=1}^N\Delta_i$ if
\be{lateig}
\sum_{i=1}^N\epsilon(\k_i+\k'_{\pi(i)})=
\sum_{i=1}^N\epsilon(\k_i+\k'_i)
\ee
for all $\pi\in S_N$, in which case (\ref{lateig}) provides the eigenvalue. 
The analog of Theorem 1 can be derived, based on the
following observation.
For $2d$-dimensional complex vectors $\u$ and $\v$ 
let us introduce the dot product
$\u\cdot\v=\sum_{j=1}^{2d}u_jv_j$, that is,
without taking the complex conjugate in one of the factors. Then, setting
\be{v(k)}
\v(\k)=(\cos k_1,\ldots,\cos k_d,i\sin k_1,\ldots,i\sin k_d)\ ,
\ee
the condition (\ref{lateig}) is equivalent with
\be{latcond}
\sum_{i=1}^N\v(\k_i)\cdot\v(\k'_{\pi(i)})
=\sum_{i=1}^N\v(\k_i)\cdot\v(\k'_i)\ .
\ee
The analog of the particular solution $\k_i\cdot\k'_j=0$ in the continuum
case has to satify
$\v(\k_i)\cdot\v(\k'_j)=0$ or $\sum_{l=1}^d\cos(k_{il}+k'_{jl})=0$
for all $i,j$. For $L$ odd no allowed pair can solve these equations.
For $L$ even, in two dimensions, a family of solutions is obtained by choosing
%any complex $\vartheta$ and
$k_{i2}=\pi/2-k_{i1}$ and $k'_{i2}=\pi/2-k'_{i1}$. If 4 is a
divisor of $L$, the same
choice works in three dimensions if $k_{i3}+k'_{j3}=\pi/2$ for all $i,j$,
e.g. $k_{i3}=k'_{i3}=\pi/4$
for all $i$. More solutions, in particular, solutions also for $L$ odd
are provided by the following theorem.

{\bf Theorem 3}
{\it Let $\K$, $\K'$ be an allowed pair.
The following assertions are equivalent.\\
(i) Equation (\ref{latcond}) holds for the permutations listed in part (i)
of Theorem 1.\\
(ii)
\be{lator}
[\v(\k_{i+1})-\v(\k_i)]\cdot[\v(\k'_{j+1})-\v(\k'_j)]=0\quad,\quad
1\leq i,j\leq N-1\ .
\ee
(iii) Equation (\ref{latcond}) holds for all permutations. As a consequence,
$\psi_{\K,\K'}$ is a (P1-3) eigenfunction of the kinetic energy and, thus,
of all Bose-Hubbard Hamiltonians with purely on-site interactions. The
eigenvalue is given by equation (\ref{lateig}).}

The proof is exactly the same as that of Theorem 1; it suffices to replace
everywhere $\k_i$ by $\v(\k_i)$ and $\k'_i$ by $\v(\k'_i)$. 
Also, condition (\ref{lator}) expresses energy conservation in two-particle
collisions, and is the same as equation (\ref{encons}), if we use the suitable
expression for the one-particle energy.

To find solutions of (\ref{lator}), it is convenient to rewrite it as
\be{lator2}
\sum_{l=1}^d[\cos(k_{i+1\,l}+k'_{j+1\,l})+\cos(k_{il}+k'_{jl})
            -\cos(k_{i+1\,l}+k'_{jl})-\cos(k_{il}+k'_{j+1\,l})]=0\ .
\ee
We will consider two families of solutions. Type 1 solutions exist for any $L$.
In two dimensions they are obtained by choosing an allowed
pair with $k_{i1}$ and $k'_{i2}$ independent of $i$. In three dimensions
we can choose e.g. $k_{i1}$, $k_{i2}$ and $k'_{i3}$ independent of $i$.
In these examples equation (\ref{lator2}) holds by a separate cancellation
of the $d$ terms. Also, notice that the pair $\K,\K'$ satisfies the equations
(\ref{orthog}). Rewriting the solution in the canonical form (\ref{mom}) or (\ref{canon}),
the constant components of the transformed vectors vanish and
$\k_i\cdot\k'_j=0$ for all $i,j$. Recalling remark 8 of the former section, this
implies that $\psi_{\K,\K'}(\R)=0$ whenever for some $i,j$ the vector
$\r_i-\r_j$ is parallel to a coordinate axis. Thus we obtain:

{\bf Corollary}
{\it Type 1 solutions of equation  (\ref{lator}) are eigenstates of the more
general lattice Hamiltonian (\ref{Hlattice}).}

Type 2 solutions exist only for $L$ even and are given by $k_{i2}=\vartheta-k_{i1}$
and $k'_{i2}=\pi-\vartheta-k'_{i1}$ which, in three
dimesions, is completed with e.g. $k_{i3}$ chosen to be independent of $i$.
Since the choice of $\vartheta$ does not influence the solution, we can
fix $\vartheta=\pi/2$, as earlier.

The solutions of (\ref{lator2}) we have presented above exist only for
$N\leq L$. Indeed, in all these examples either in $\K$ or in $\K'$ a single
component specifies the vector. Since in both sets all the vectors
must be incongruent, the defining component must take on $N$ different
values out of $L$ possible ones. In higher dimensions the same kind of
solutions can easily be given -- for example, in four dimensions we can
fix the first two components in the $\K$ set and the second two in the
$\K'$ set. In general, we can freely choose $N$ incongruent vectors from a
$[d/2]$ dimensional subspace, which limits the number of particles to $N\leq L^{[d/2]}$.
If $N\ll L^{[d/2]}$, it is possible to choose $\K,\K'$ such that
$|\k_i|,|\k'_i|\ll 1$ for each $i$. Then, in (\ref{lateig}) we can
expand the cosine functions up to second order, and find that the smallest
attainable energy is
of order $L^{-2}N^{1+{2\over[d/2]}}$.
Thus, for $N=o(L^{2[d/2]\over 2+[d/2]})$ there
are eigenstates whose energy tends to zero as $L$ goes to infinity.
They correspond to gapless excitations above the grand canonical ground state
when this is the vacuum ($N=0$ or parallel spins).

The reader may notice that part (iii) of Theorem 3 claims less than that of
Theorem 1. It is because equation (\ref{lator}) does not imply that 
$\k_i+\k'_j$ are different for different pairs $(i,j)$. 
This nevertheless follows from (\ref{orthog}) for type 1
solutions, but a counterexample of type 2 is
easily obtained. Let $L$ be even and choose $k_{i2}=\pi/2-k_{i1}$,
($k_{i3}$ independent of $i$ if $d=3$) and $\k'_i=\k_i$. As
a special choice of type 2, the pair $\K,\K'=\K$
solves equation (\ref{lator2}) and, for any $(i,j)$, $\k_i+\k'_j=\k_j+\k'_i$.
The same example shows that
in the sum over the $N!$ symmetrized plane waves yielding the (P1-3) eigenstate
$\psi_{\K,\K}$ (cf. equation (\ref{psikk'})) there are coinciding terms:
$\psi^+_{\K+\alpha\K'}=\psi^+_{\alpha^{-1}\K+\K'}$ implies that
$\psi^+_{\K+\alpha\K}=\psi^+_{\K+\beta\K}$ if
$\beta=\alpha^{-1}$. Now $\alpha$ differs
from its inverse if and only if it contains at least one cycle
longer than two. Thus, our example works for
all such $\alpha$ (but we need $N>2$) and $\beta=\alpha^{-1}$.
As a permutation and its inverse
have the same parity, the two occurrences of the same symmetrized plane wave arrive with
the same sign in $\psi_{\K,\K}$.

Keeping the earlier definition of minimality, Lemma 2 remains valid: Minimal functions
form a basis among the (P1-3) eigenstates of $-\sum\Delta_i$. Type 1 eigenfunctions
are minimal, as it is
seen by applying Lemma 4. We believe that all solutions of equations (\ref{lator}), and
only them, are minimal functions. However,
we cannot prove this and have only a weaker analog of Theorem 2.

{\bf Theorem 4}
{\it If $\psi$ is a minimal eigenfunction of $-\sum\Delta_i$ then
${\rm supp}\,\psi\subset\{\psi^+_{\K+\alpha\K'}\}_{\alpha\in S_N}$ for some allowed pair
$\K,\K'$ satisfying equations (\ref{lator}).}

For $N=2$ the assertion is equivalent with the stronger claim of Theorem 2.
The second of equations (\ref{pair}) is to be replaced by
\be{pairlat}
\sum_{j=1}^d(\cos p_{1j}+\cos p_{2j})=\sum_{j=1}^d(\cos q_{1j}+\cos q_{2j})\ .
\ee
Again, the trivial solution $\{\p_1,\p_2\}=\{\q_1,\q_2\}$ is to be excluded,
but now the four vectors need not be different. Indeed, with our
earlier example $k_{i2}=\pi/2-k_{i1}$ and $\k'_i=\k_i$ for $i=1,2$,
$\p_i=2\k_i$ and $\q_1=\q_2=\k_1+\k_2$ is a nontrivial solution.
When parametrizing
a solution (\ref{N=2}) with an allowed pair $\K,\K'$, equation (\ref{pairlat})
coincides with (\ref{lator}) written for $i=j=1$. The minimality
of such a solution is trivial, and the proof that no other minimal function exists is
the same as in Theorem 2. The weaker assertion for a general $N$ can be shown as in Theorem 2
by starting with a weaker induction hypothesis.
\vspace{4pt}\\
{\bf Acknowledgment}\\
The author thanks L\'aszl\'o Erd\H os, G\'abor F\'ath, Philippe Martin and
Valentin Zagrebnov for discussions, comments and criticism.
This work was supported by Grant T 30543 of the Hungarian Scientific Research Fund.

\end{document}